\documentclass{vienna-conf2019}
\usepackage{graphicx}
\usepackage{hyperref}
\usepackage[]{natbib}  
\usepackage{epstopdf}

\def\BibTeX{{\rm B\kern-.05em{\sc i\kern-.025em b}\kern-.08em
    T\kern-.1667em\lower.7ex\hbox{E}\kern-.125emX}}
\bibpunct{(}{)}{;}{a}{}{,}  %%%%%%%%%%%%%  A&A bibliography style
\begin{document}

\TitreGlobal{Stars and their variability observed from space}

%%-----------------------------------------------------------------
%%      the top matter
%%

\title{R-mode oscillations in eclipsing binaries}

\runningtitle{R-mode oscillations}

\author{Hideyuki Saio}\address{Astronomical Institute, Tohoku University, Sendai, Japan}

%% Keep this line, even if the page will be settled afterwards.
\setcounter{page}{1}
%\setcounter{page}{237}

%%-----------------------------------------------------------------

\maketitle

%%-----------------------------------------------------------------
%%        The abstract
%% 
%%  Warning!  within the abstract:
%%  - do not use macros. 
%%  - do not use commands like: \cite, \citet, \citep ... etc.

\begin{abstract}
The presence of r mode oscillations (global modes of Rossby waves coupled with buoyancy) in an eclipsing binary is recognized as frequency groups in a frequency-amplitude diagram obtained from a Fourier analysis of Kepler light curve data.  The frequency at the upper bound of an r-mode frequency group is close to the rotation frequency of the star. Analysing about eight hundred Kepler light curves of eclipsing binaries finds about seven hundred cases showing the signature of r modes. Sometimes two sets of the frequency groups are found, which indicates the two component stars to have slightly different rotation frequencies. Plotting thus obtained rotation frequencies with respect to orbital frequencies, we find that rotation is roughly synchronous to the orbital frequency if the latter is larger than about 1~c/d, while some stars rotate super-synchronously in systems with longer orbital periods.  
\end{abstract}


%%-----------------------------------------------------------------

\section{Introduction}
Possible presence of r mode oscillations (global modes of Rossby waves coupled with buoyancy; sometimes called `rossby modes') in rotating stars has been discussed theoretically for long \citep[e.g.,][]{pap78,pro81,sai82}.
But the first clear evidence of r modes was discovered only recently by \citet{vanr16} in Kepler data of rapidly rotating $\gamma$ Dor variables; the evidence of r modes is the period-spacing increasing with period, while the period spacing of prograde g modes decreases with period.
Later, from the period-spacing property \citet{LiG19_rg, LiG19_600} found r modes in a large fraction of rapidly rotating $\gamma$ Dor stars.

Another property of r modes is a frequency group located just below the rotation frequency in frequency-amplitude diagrams, which is useful to identify r modes when each frequency in the group cannot be resolved. 
From this property \citet{sai18,sai18c} found r modes in Kepler data for spotted early-type stars, binary stars, chemically-peculiar stars, and Be stars. 
\citet{sai19} discusses that short period oscillations observed in accreting white dwarfs in cataclysmic variables are consistent with r mode oscillations.
We note that this possibility was already discussed more than 40-years ago by \citet{pap78}.
  
Here, we discuss r mode oscillations in eclipsing binaries adopting Kepler data from the Kepler Eclipsing Binary Catalog (KEBC) V3 maintained by the eclipsing binary working group \citep[e.g.][\url{http://keplerebs.villanova.edu/}]{kebc16,kebc12,kebc2,kebc1}. 
We identify r modes from the property of frequency groups in frequency-amplitude diagrams, which will be discussed below.

\section{Basic properties of r modes}
%%-------------------------

\begin{figure}[ht!]
 \centering
 \includegraphics[width=0.48\textwidth,clip]{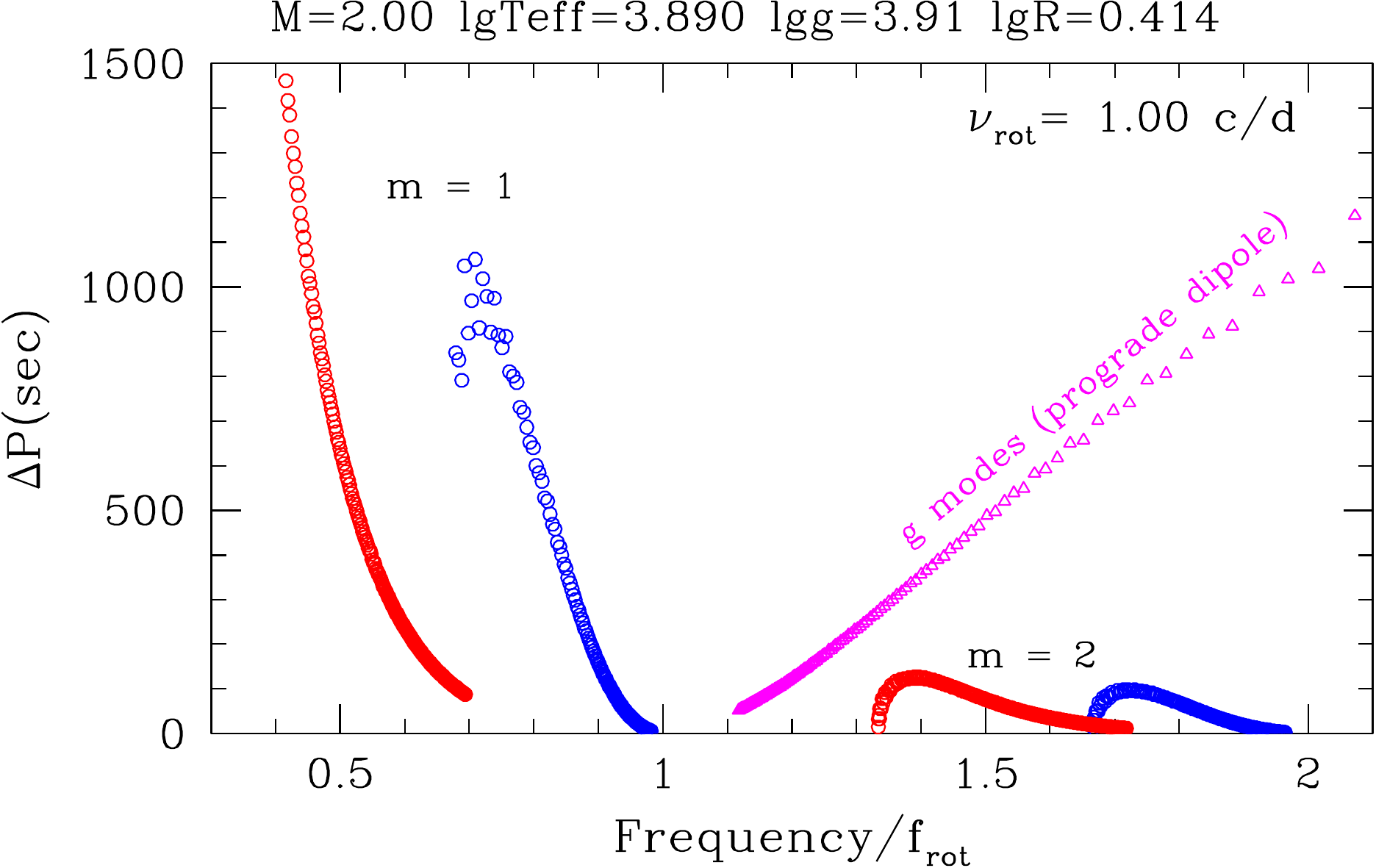} %%{freq_dp_rg_modes_crop}      
\hspace{0.02\textwidth}
 \includegraphics[width=0.48\textwidth,clip]{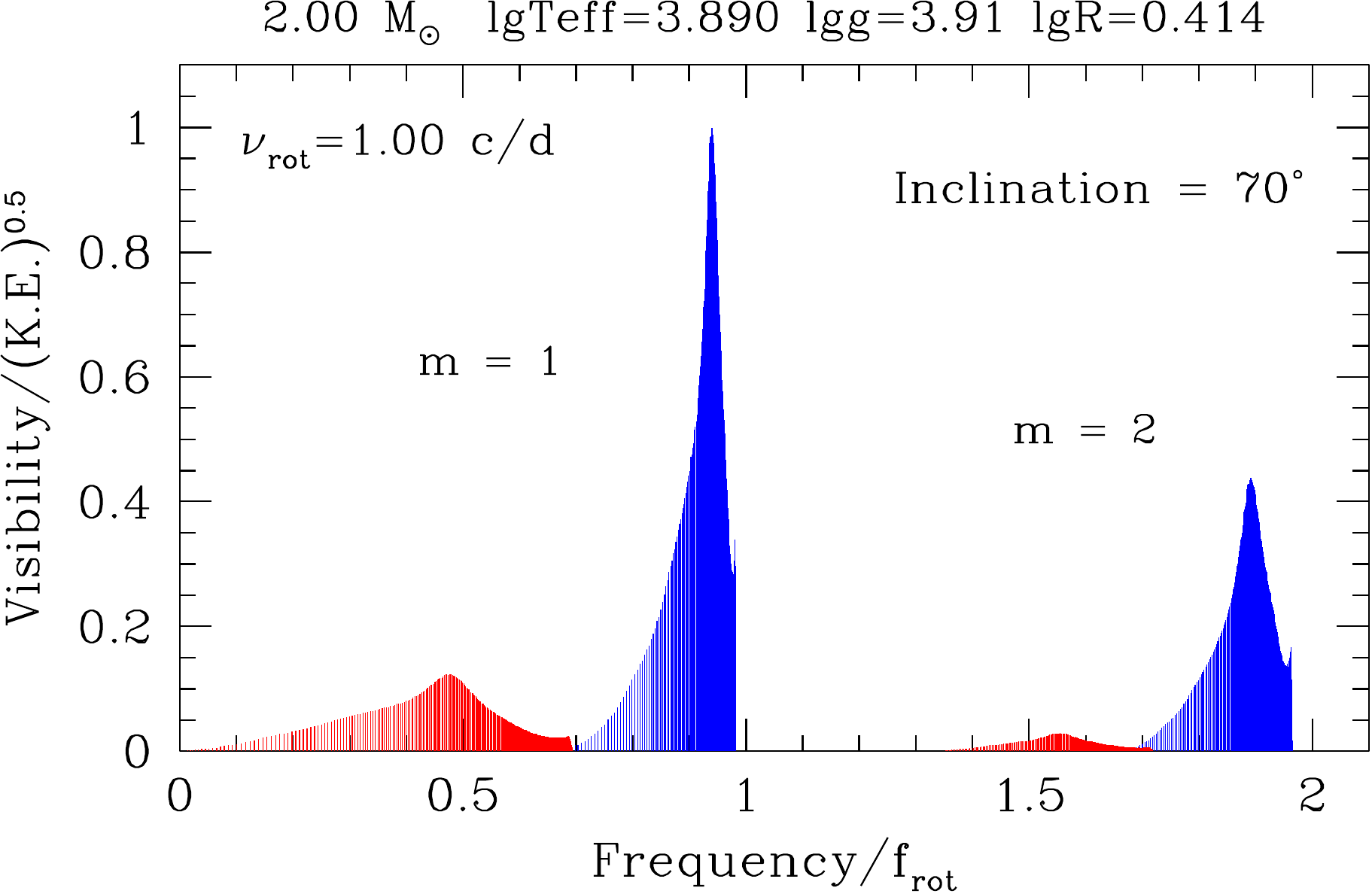}%%{visibility_m1m2_crop}      
 
  \caption{{\bf Left:}Period spacings versus frequency of r modes (and dipole prograde g modes (magenta triangles) for comparison) in the inertial frame.  Although period spacing is usually plotted as a function of period, here it is given as a function of frequency for better correspondence to the diagram in the right side. Blue and red colors are used for even and odd (with respect to the equator) r modes. {\bf Right:} Visibility distribution of r modes (arbitrarily normalized) expected for energy equipartition as a function of frequency in the inertial frame.}
  \label{fig:rmode}
\end{figure}

R-modes are normal modes of large scale Rossby waves, which have nearly toroidal motions but couple with buoyancy due to the effect of Coriolis force.  
In the co-rotating frame, r-modes are retrograde modes and have frequencies smaller than the rotation frequency. 
Therefore, in the inertial frame, r modes are observed as prograde modes with phase speeds less than the rotation speed, and the frequencies of r modes with azimuthal order $m$ ($>0$; we adopt in this paper the convention that $m>0$ for retrograde modes in the co-rotating frame) are less than $m\nu_{\rm rot}$ with $\nu_{\rm rot}$ being cyclic frequency of rotation. Period spacing in the co-rotating frame is roughly constant \citep[see e.g.][for details]{sai18}, while in the inertial frame, period spacing, $\Delta P$, decreases with frequency (Fig.~\ref{fig:rmode}), or increases with period.
The sign of the slope of $\Delta P$ with respect to frequency (or period) of r modes in the inertial frame is opposite to that of g modes. From this fact r modes can be detected convincingly if each r-mode frequency is resolved as in many amplitude-period diagrams obtained from Kepler data for $\gamma$ Dor stars \citep{vanr16,LiG19_rg,LiG19_600}.

For other type variables, however, each r-mode frequency is not resolved, but r modes appear as dense frequency groups in frequency-amplitude diagrams. As seen in the right panel of Fig.~\ref{fig:rmode}, even (i.e., symmetric with respect to the equator) r modes, in particular $m=1$ mode-group, are most visible, whose frequencies are slightly lower than $m\nu_{\rm rot}$. In other words, if we find an r-mode frequency group of $m=1$, we can estimate an approximate rotation frequency as the frequency at the upper bound of the group.    

\section{R modes in eclipsing binaries}
%%---------------------------------

\begin{figure}[ht!]
 \centering
 \includegraphics[width=0.48\textwidth,clip]{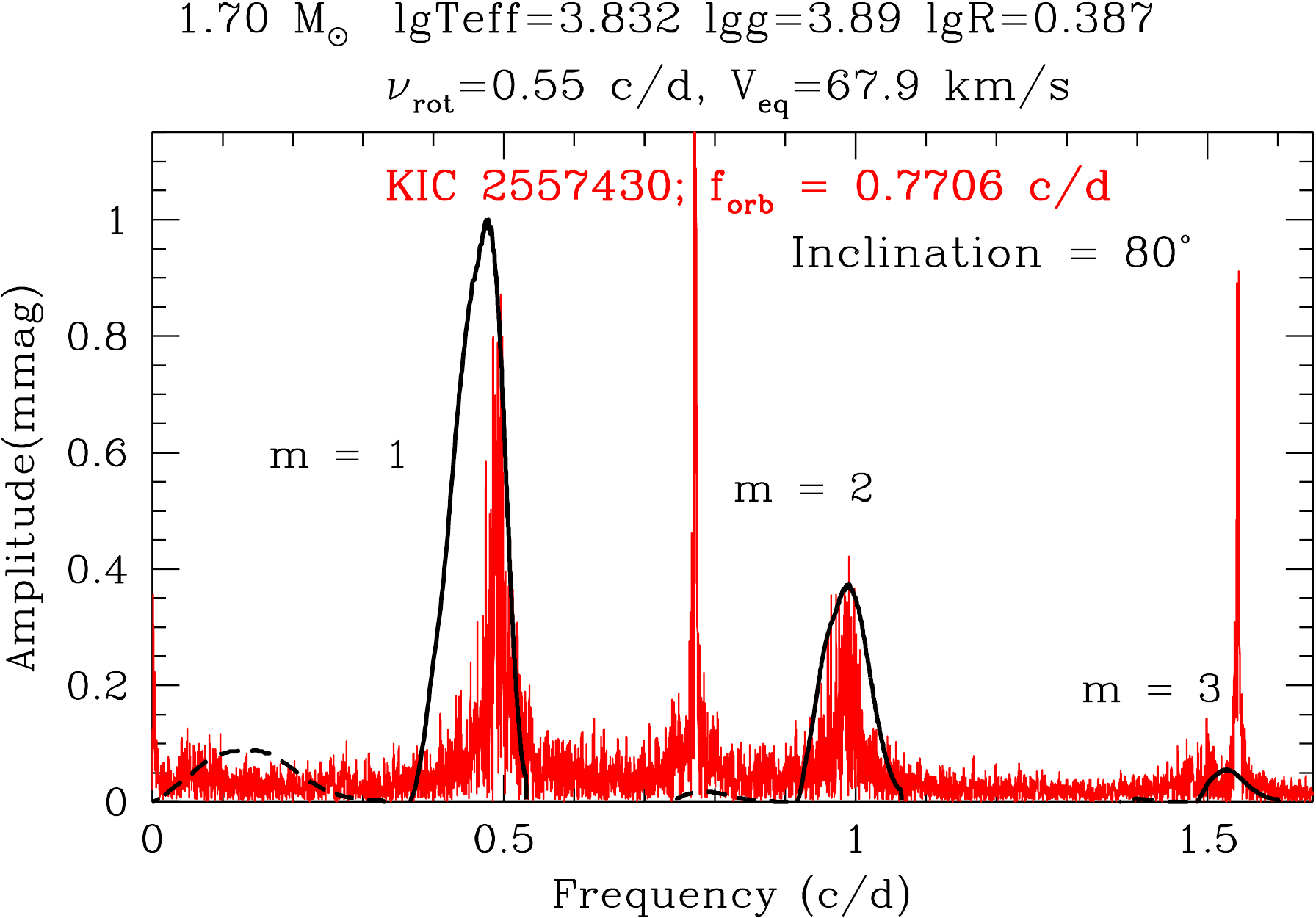}%%{m1p70_15_0p55cd_k2557430_crop}%      
\hspace{0.02\textwidth}
 \includegraphics[width=0.48\textwidth,clip]{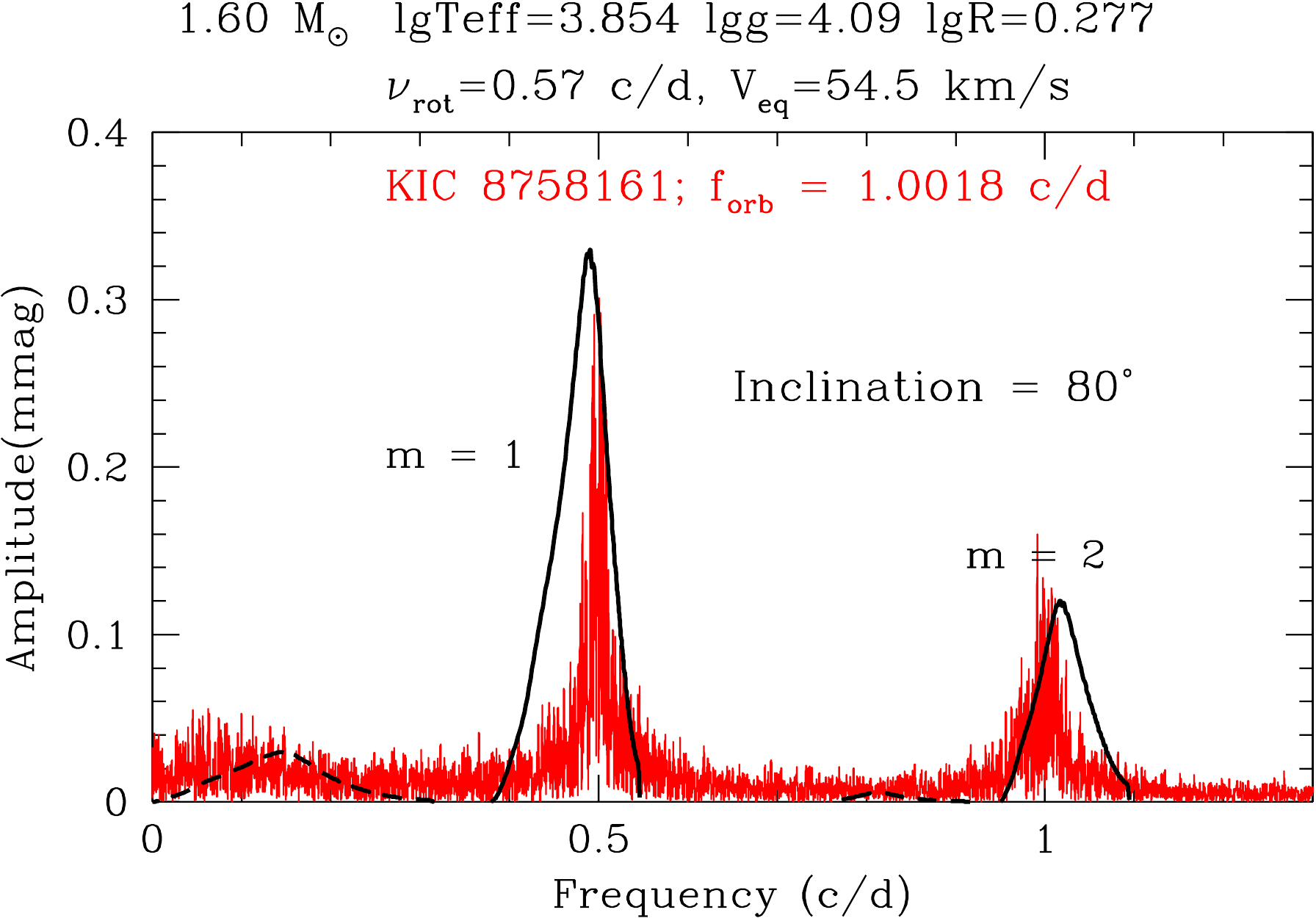}%%{m1p60_10_0p57cd_k8758161_crop}      
  \caption{{\bf Left:} Frequency-amplitude relation (red lines) of the eclipsing binary KIC~2557430 with an orbital frequency of 0.771 c/d is fitted with theoretical r-mode visibility predicted for a $1.70~M_\odot$ model at a rotation frequency of 0.55~c/d, where solid and dashed lines are for even and odd modes, respectively. {\bf Right:} Similar to the left diagram but for KIC~8758161 with an orbital frequency of 1.00 c/d fitted with theoretical r-mode visibility of a $1.6~M_\odot$ model at a rotation frequency of 0.57 c/d. }
  \label{fig:ex1}
\end{figure}

From the Kepler Eclipsing Binary Catalog V3 (KEBC), 837 binaries have been chosen in a range of orbital frequencies from 0.2 to 2.5~c/d, under the condition that they were observed by Kepler longer than 3 quarters, and the primary's effective temperatures are higher than 4000~K. 

\begin{figure}[ht!]
 \centering
 \includegraphics[width=0.48\textwidth,clip]{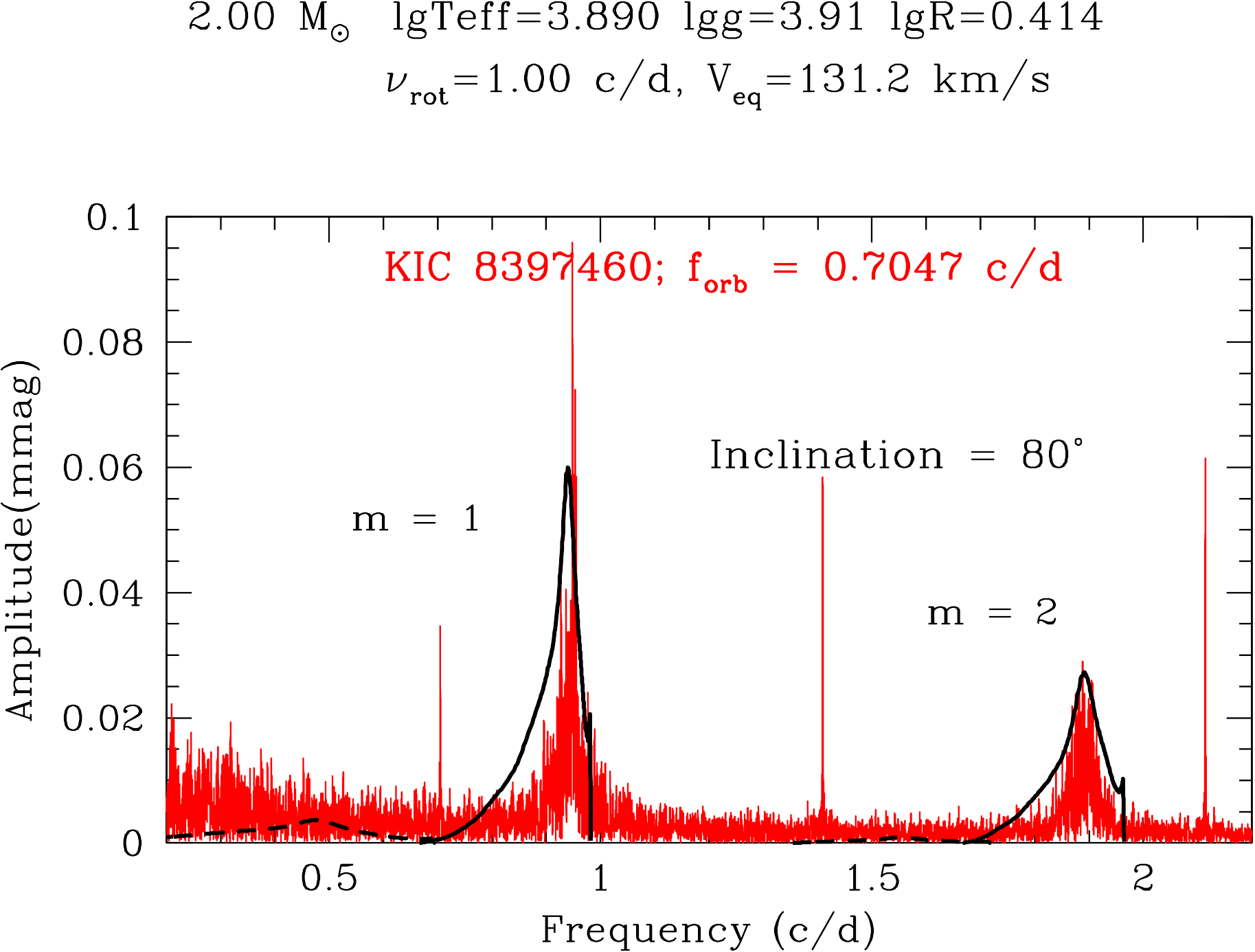}%%{m2p0_13_1p00cd_k8397460_crop}%
 \hspace{0.02\textwidth}      
 \includegraphics[width=0.48\textwidth,clip]{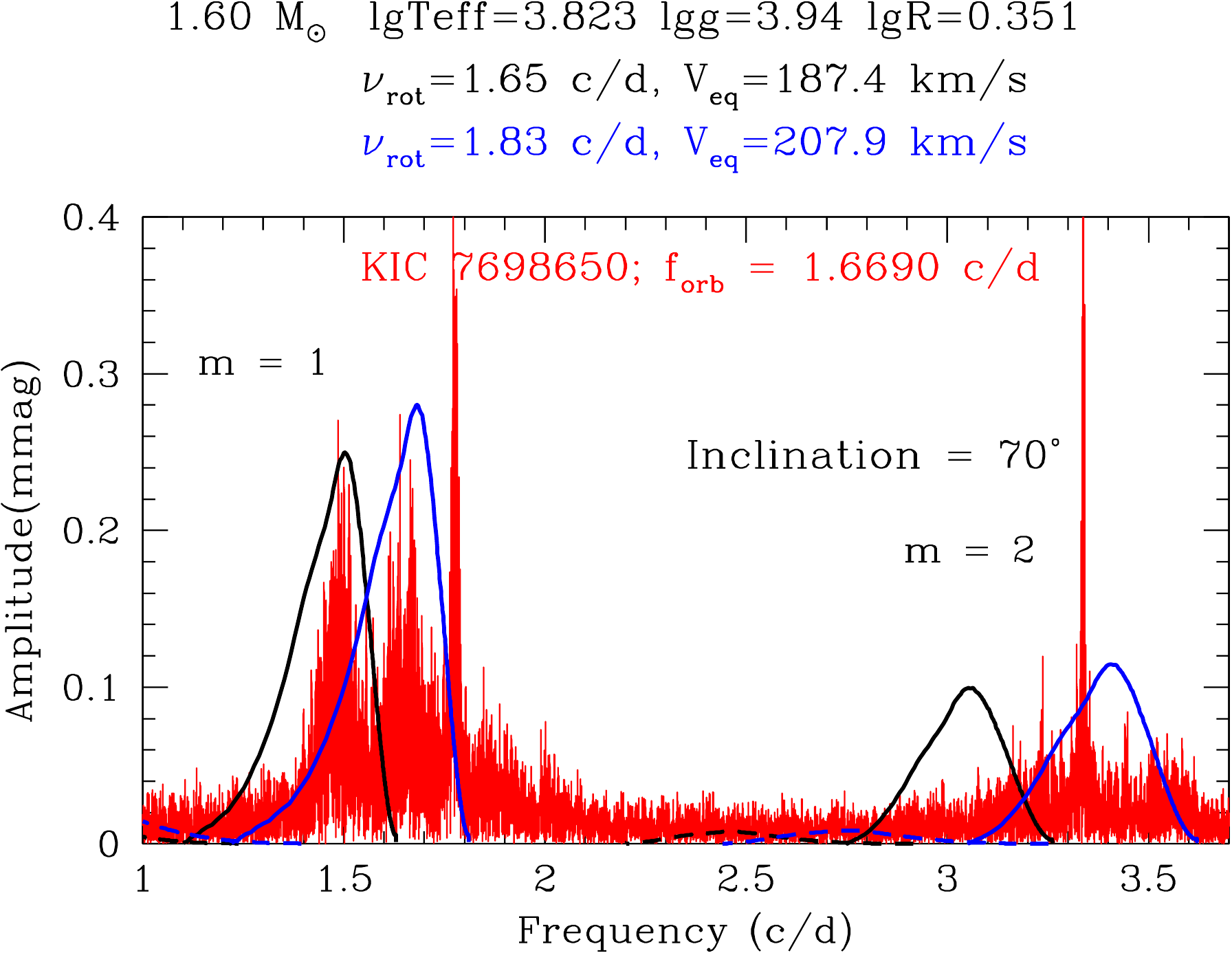}%%{m1p60_16_1p65cd_1p83cd_k7698650_crop}      
  \caption{{\bf Left:} Similar to Fig.~\ref{fig:ex1} but for KIC~8397460 with an orbital frequency of 0.70~c/d fitted with theoretical visibility prediction of r modes for a $2.0~M_\odot$ model at an rotation frequency of 1.00~c/d.  {\bf Right:} Similar to the left figure but for the eclipsing binary KIC~7698650 with an orbital frequency of 1.67~c/d. There are two closely-separated frequency groups around $\sim1.5$~c/d and $\sim1.7$~c/d. They are fitted with r-mode visibility models of $1.6~M_\odot$ with rotation frequencies of 1.65~c/d and 1.83~c/d, respectively. These two frequency groups correspond to r mode oscillations in the two components of the binary.}
  \label{fig:ex2}
\end{figure}

For each case, mean light curve of eclipses is subtracted from the original Kepler light-curve data using the polynomial fit given in the KEBC.
A Fourier analysis by the software PERIOD04 \citep{period04} for the residual data yields a frequency-amplitude diagram, which is
 searched by eye for frequency groups attributable to r modes.
R-mode features are found in 737 binaries, and among them, 320 cases show r-mode features from both components.   
Figs.~\ref{fig:ex1}~and~\ref{fig:ex2} show four examples of frequency-amplitude diagrams and fits with theoretical visibility distributions of r modes (solid and dashed lines) derived from models with best-fit rotation frequencies, where model parameters were adopted by referring to the effective temperatures obtained by \cite{arm14}.
We note that the frequency-amplitude diagram in the right panel of Fig.~\ref{fig:ex2} is an example showing r modes from both two component stars rotating at slightly different rates from each other. 

\section{Rotation frequencies of stars in eclipsing binaries}
%%---------------------
\begin{figure}[ht!]
 \centering
 \includegraphics[width=0.48\textwidth,clip]{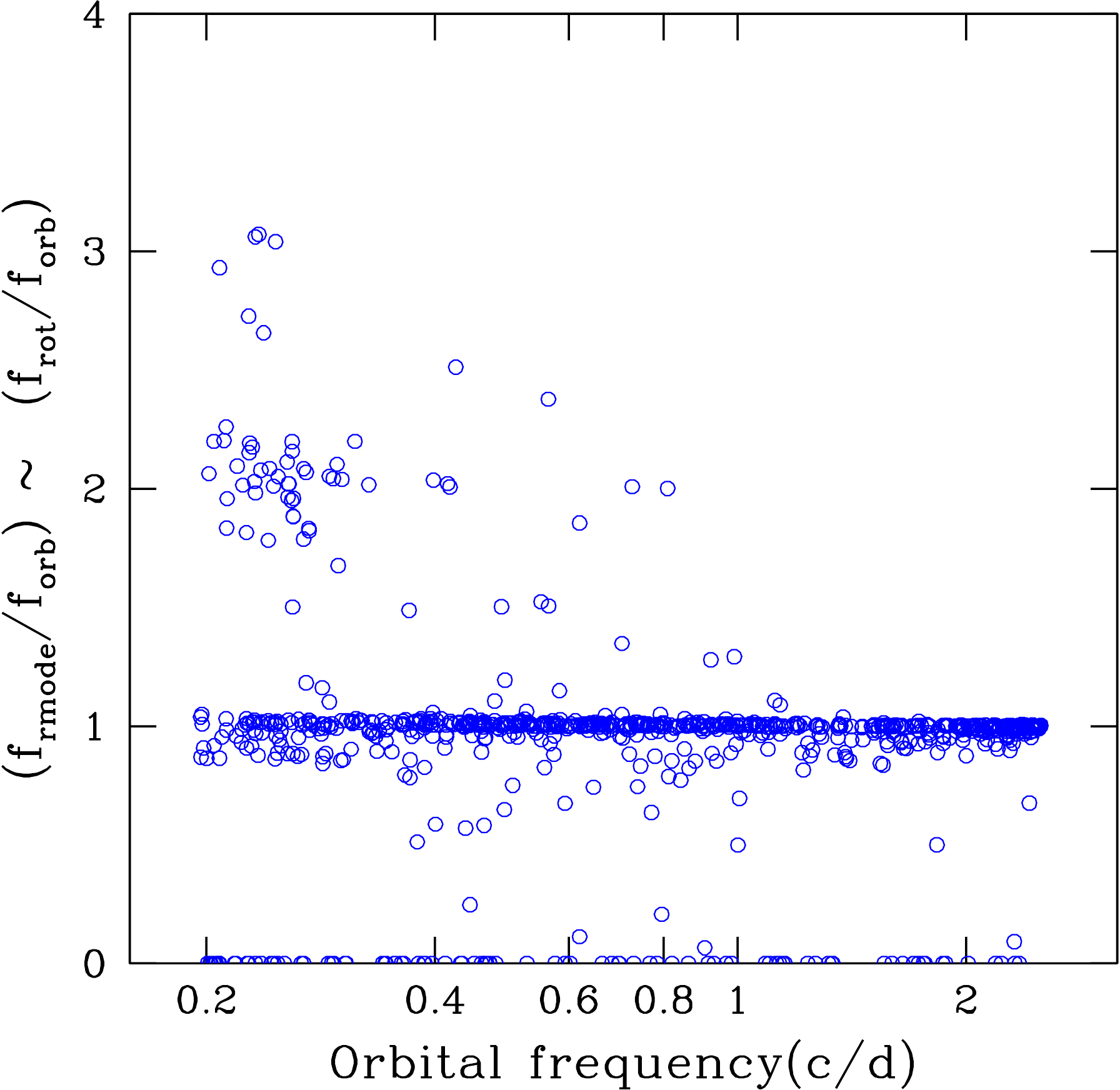}%%{forb_frot_primary_crop}% 
 \hspace{0.02\textwidth}     
 \includegraphics[width=0.49\textwidth,clip]{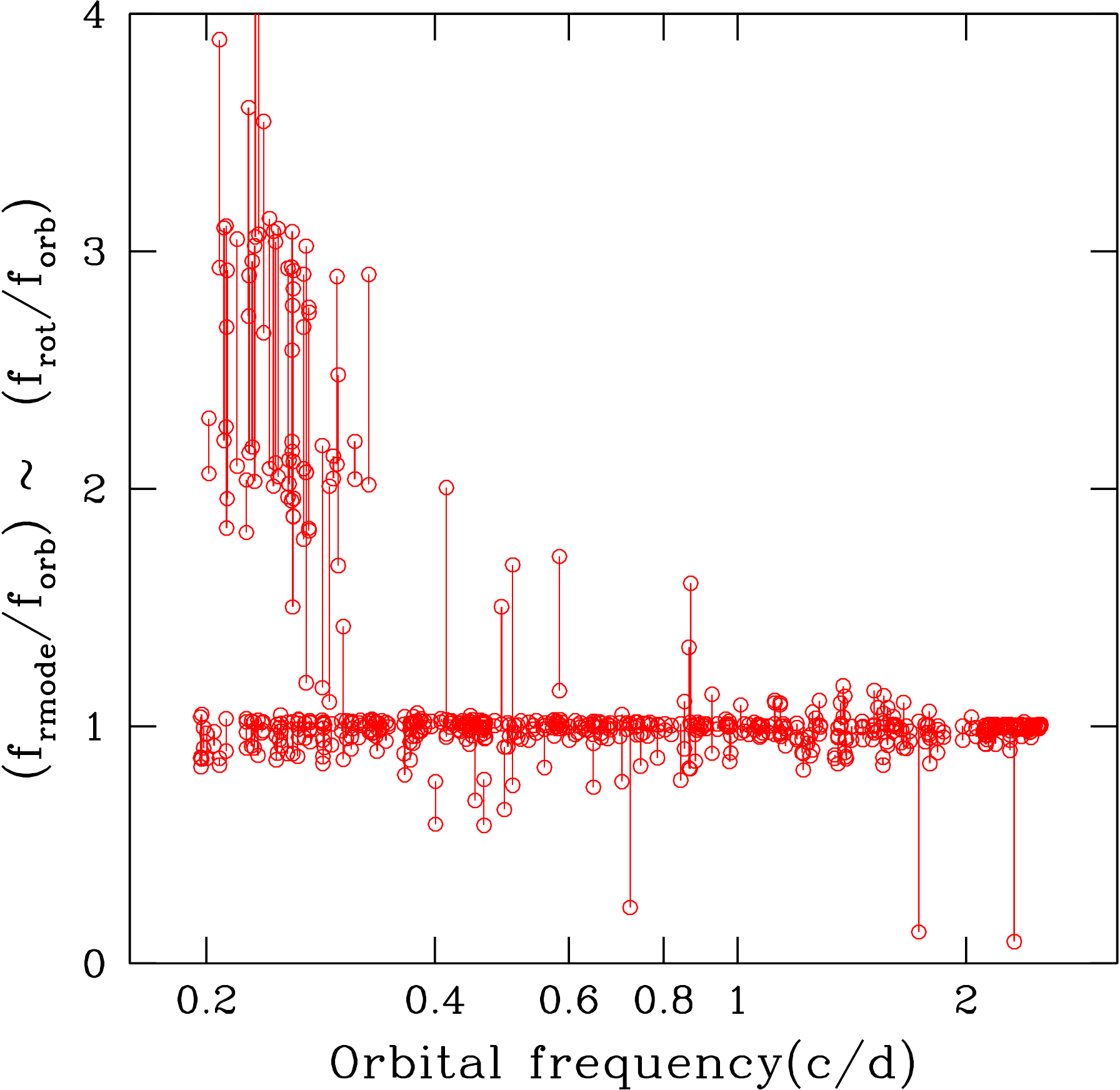}%%{forb_frot_double_crop}      
  \caption{{\bf Left:} Frequencies at upper bounds of $m=1$ r-mode frequency groups, which should be close to rotation frequencies, relative to orbital frequencies for eclipsing binaries selected from the KEBC. Points on the horizontal axis are cases in which no r modes are detected.
  {\bf Right:} The same as the left panel but for the cases in which r-mode frequencies of both components (connected by vertical lines) are detected. }
  \label{fig:rot}
\end{figure}

In the left panel of Fig.~\ref{fig:rot}, an open circle gives the ratio of the frequency at the upper bound of the r-mode frequency group (, which should be similar to the rotation frequency) to the orbital frequency of an eclipsing binary. Non-detection cases lie on the horizontal axis.
This figure indicates that stars in eclipsing binaries with orbital frequencies larger than about 1~c/d tend to rotate more-or-less synchronously to the orbital motion, while in some longer-period binaries rotation frequencies are considerably higher than orbital frequencies.

The right panel of Fig.~\ref{fig:rot} is the same as the left panel but for the eclipsing binaries which show two sets of closely separated r mode frequency groups which should correspond to r modes in each component star.
If the orbital frequency is less than about 1~c/d, rotation frequencies of the two component stars sometimes differ considerably from each other.

\section{Conclusions}
%%--------------------
We found that r-mode oscillations are present frequently in eclipsing binaries (737 cases are found among 837 samples), which suggests that fluid motions arisen by  tidal force should generate Rossby waves and hence r-mode oscillations.
Detecting r-mode frequency groups is useful to obtain a rotation frequency of a star or sometimes rotation frequencies of both components in eclipsing binaries.
These rotation frequencies relative to orbital frequencies would give insight about the orbit-rotation interaction in close-binary stars.
  
 Although the results presented in this paper are still preliminary, they seem to indicate that in the binary systems whose orbital periods are shorter than about one day, stellar rotation is more-or-less synchronized, while in the systems with longer orbital periods, stellar rotation periods can be considerably shorter than  the orbital periods. Further accumulation of analyses is needed to have a clearer picture on the orbit-rotation interactions in close binaries.

% -------------------------
\begin{acknowledgements}
The author is very grateful to the Kepler Eclipsing Binary working group for their maintenance of the Kepler Eclipsing Binary Catalog.
\end{acknowledgements}

%%-----------------------------
%%   Bibliography
%%-----------------------------
%%
%% The format for references is the one adopted by A&A (see the example below).
%% To set the reference list in the proper format, we ask you to use BibTEX and %% the natbib package instead of the standard 'thebibliography' environment.
%%
%% The following lines are then required:
\bibliographystyle{aa}  % A&A bibliography style file (aa.bst)
\bibliography{saio_6o01} % your references in file: surname_nxnn.bib

\end{document}